\documentclass{appolb}
\usepackage{epsfig}

\begin{document}
\title{SHELL MODEL STUDY OF EVEN-EVEN $^{132-136}$Te
\\NEUTRON-RICH NUCLEI
}
\author{F. A. Majeed
\address{ Department of Physics, College of Science, Al-Nahrain University,\\ Baghdad, Iraq
\and The Abdul Salam International Centre for Theoretical Physics}
}
 \maketitle
\begin{abstract}
Large-scale shell model calculations were performed for neutron-rich
even-even $^{132-136}$Te using a realistic effective interaction
derived from CD-Bonn nucleon-nucleon potential for the positive and
negative parity states. The calculated results are compared with the
recently available experimental data and with the recent theoretical
work. The transition rates $B$($E$2; 0$^+\rightarrow$2$^+$) are also
calculated by taking into consideration core polarization effect by
choosing best effective charges for proton and neutron. The result
of our theoretical calculations are compared with experimental data
and with the previous theoretical work. A very good agreement were
obtained for all nuclei.

\end{abstract}

\section{Introduction}
In recent years, there has been significant progress in the
experimental knowledge of neutron-rich nuclei with few particles or
holes outside doubly magic $^{132}$Sn \cite{MM02}. The region of
neutron-rich Sn and Te nuclei with the number of protons at, or just
above, the Z=50 closed shell becomes recently important for study
both experimentally and theoretically \cite{BA05}. Previous
shell-model calculations (see Ref.~\cite{rad02} for details)
provided reasonable agreement with energy spectra and $B$($E$2) in
N=80 and N=82 Sn and Te isotones but failed to explain the $B$($E$2)
value in $^{136}$Te. Magnetic moments were calculated for
$^{134}$Te, $^{136,137}$Xe and $^{137}$Cs by Sarkar and Sarkar
\cite{sar01} with the KH5082 and CW5082 interactions fitted in the
$^{208}$Pb and scaled to the $^{132}$Sn region, and with empirical
effective single-particle $g$-factors. Shell-model calculations for
the 2$^+$, 4$^+$ and 6$^+$ states in $^{130-134}$Te and
$^{132-136}$Xe were reported in Ref.~\cite{jak02}, where the surface
delta interaction (SDI) was used with two different sets of
parameters. The single particle states were chosen to reproduce
single proton states in ${^{133}_{51}}$Sb and single neutron states
in ${^{131}_{50}}$Sn. The single-particle spin and orbital effective
$g$-factors were based on the experimental $g$-factors of the
low-lying (7/2)$^+$ and (5/2)$^+$ states in the odd-$Z$, $N=82$
isotones. Shell model calculations with modified empirical
hamiltonian that obtained by some modifications of a Hamiltonian
(CW5082) originally derived from the $^{208}$Pb region and scaled to
the $^{132}$Sn region were performed by Sarkar and Sarkar
\cite{suk04}. The $g$-factor of the first 2$^+$ state in $^{132}$Te
and energy levels of nuclei near $^{132}$Sn has been studied by
Brown {\em et al.}\cite{BA05} using microscopic interaction based on
CD-Bonn nucleon-nucleon interaction \cite{rmf96}.

The aim of the present work is to study the level energies including
the high $J$${^\pi}$-values, which is not studied before to test the
ability of the "new developed effective interaction obtained
starting with a $G$ matrix derived from CD-Bonn nucleon-nucleon
interaction by Brown {\em et al.} \cite{BA05} and this interaction
codenamed lately as SN100PN in the new released version of Oxbash
for windows \cite{oxb04}" in reproducing the $J$${^\pi}$-values. The
main difference from the present work and Brown {\em et al.} work is
that we consider the core to be $^{100}$Sn instead of $^{132}$Sn
thats means our choice of core as $^{100}$Sn, the valence neutrons
are particles not not like the case of Ref.\cite{BA05} were the
valence neutrons are holes for the case of N $\leq$ 82, trying to
investigate if this change of choosing different core would enhance
the calculations of level spectra for $^{132}$Te and $^{134}$Te. On
the other hand, since the electromagnetic transition rates provide
one of the most sensitive probes of nuclear structure, therefore it
is studied in this work for first 2${^+}$ , to give clear picture of
the present large-scale shell model calculations in reproducing the
experiment.
\section{Outline of calculations}
In our calculations the core is considered as $^{100}$Sn with 32 and
34 particles outside the core for $^{132}$Te and $^{134}$Te
respectively. The model space SN100PN were used with SN100PN
hamiltonian \cite{BA05} based on CD-Bonn renormalized $G$ matrix
using the code OXBASH \cite{oxb04}. The single particle energies
that used in the present work are quoted from Ref. \cite{BA05} for N
$\leq$ 82 as follows, the proton single-particle energies are
$-$9.68, $-$8.72, $-$7.24, $-$7.34
 and $-$6.88 MeV for the proton model space
$  0g_{7/2}  $, $ 1d_{5/2} $ ,  $ 1d_{3/2} $ , $  2s_{1/2}  $ and $
0h_{11/2} $, respectively. The neutron single-particle energies are
$-$9.74, $-$8.97, $-$7.31, $-$7.62  and $-$7.38 MeV for the neutron
model space $ 0g_{7/2}  $, $ 1d_{5/2} $  ,   $ 1d_{3/2} $ , $
2s_{1/2} $ and $ 0h_{11/2} $, respectively.

For $^{136}$Te where N $\geq$ 82 the core is taken as $^{132}$Sn
with 4 particles outside the core. The effective interaction KH5082
\cite{kuo92} which is originally fitted for $^{208}$Pb and scaled to
the $^{132}$Sn region were employed in the calculations of
$^{136}$Te by changing the single-particle energies (SP) used with
KH5082 effective interaction by those quoted from Ref.\cite{BA05} as
follows, the proton single particle energies were taken for the same
model space for protons as above, while the neutron orbits should be
changed by those quoted from Ref. \cite{BA05} ad follows,
$-$$0.894$, $-$$2.455$, $-$$0.450$, $-$$1.601$, $-$$0.799$ and
$0.25$ MeV for the neutron model space $0h_{7/2}$, $1f_{7/2}$,
$1f_{5/2}$, $2p_{3/2}$, $2p_{1/2}$, $0i_{13/2}$, respectively.

\section{Results and comparison with experiment}
\subsection{Excitation energies}
The calculated excitation energies for $^{132}$Te are presented in
Fig.1. Good agreement were obtained by comparing our theoretical
calculations with the experiment for both positive and negative
parity states. Our calculations are very close and sometimes exactly
as the the results obtained from Ref.\cite{BA05} and if we focus our
attention to the predication of the first 2$^+$ our work predicts
this level at 954 keV and the experiment is 974 keV the absolute
difference between the two values is 20 keV and if we compare this
result with the previous theoretical work of Terasaki $et al.$
\cite{HR04,ter02} they use quasi random phase approximation (QRPA)
and their theoretical work predicts 2$^+$ at 1211 keV and the
absolute difference between experiment is 237 keV our result are in
significantly better agreement with experiment.

Figure 2 presents the comparison of results obtained from this work
with the experimental data and with previous theoretical work
obtained from Ref. \cite{BA05} for $^{134}$Te. From the figure we
can notice that this model is in good agreement with experiment up
to $J \leq 8$,but it there is large discrepancy in predicting 9$^+$
and 10$^+$ in comparison with experiment this reflects the
inadequacy of the model space. If we compare the prediction of the
first 2$^+$ our work predict this level at 1211 keV and the absolute
difference with experimental value is 68 keV in comparison with the
previous theoretical work from Ref.\cite{Gar04}, their work predict
this level at 1375 keV with absolute difference 96 keV from
experimental value.

In Fig.3, all the experimental and calculated levels up to 7782 keV
with the comparison of the previous theoretical work taken from
Ref.\cite{suk04} are reported for $^{136}$Te. Good agreement were
obtained by comparing our theoretical calculations with the
experimental values up to $J$${^\pi}$=10$^+$, but for the higher
spin $J > 12$ the calculations starts to deviates from the
experimental values and this reflects the inadequacy of the model
space. The first 2$^+$ is predicted at 866 keV which is very close
the value predicted by S. Sarkar and M. Sarkar \cite{sar01}, but
their work \cite{suk04} using empirical hamiltonian named SMPN5082
obtained from some modifications of a Hamiltonian CW5082, their work
predicts the energy levels for positive and negative parity states
much better than this work.
\begin{figure}
\centering
\includegraphics[width=2.0 in]{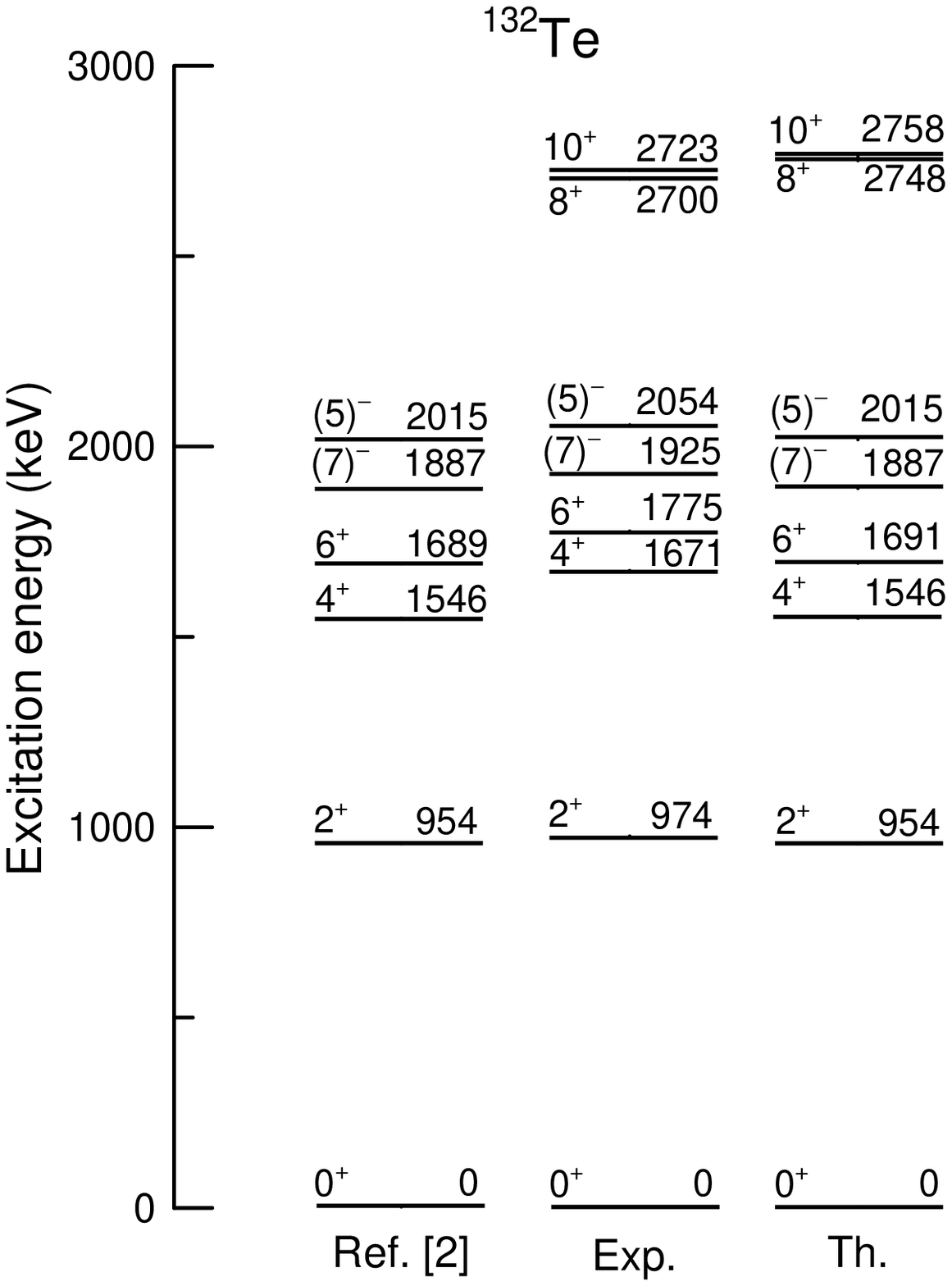}
\caption{Calculated energy levels for positive and negative parity
states of $^{132}$Te in comparison with experiment taken from
Ref.\cite{HR04} and the previous theoretical work taken from
Ref.\cite{BA05}.}
\end{figure}

\begin{figure}
\centering
\includegraphics[width=3.45 in]{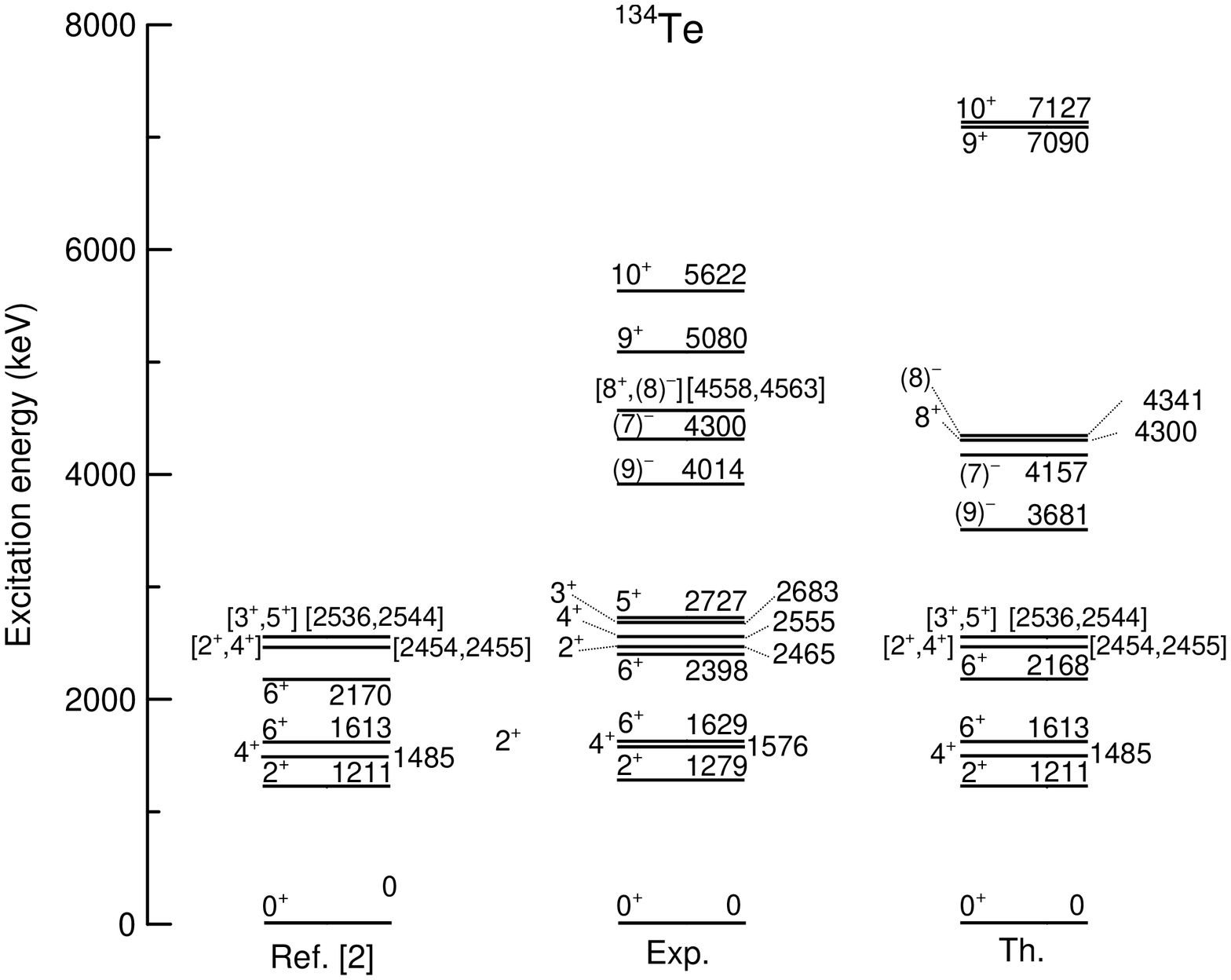}
\caption{Calculated energy levels for positive and negative parity
states of $^{134}$Te in comparison with experiment taken from
Ref.\cite{PJD97} and the previous theoretical work taken from
Ref.\cite{BA05}.}
\end{figure}

\begin{figure}
\centering
\includegraphics[width=2.6 in]{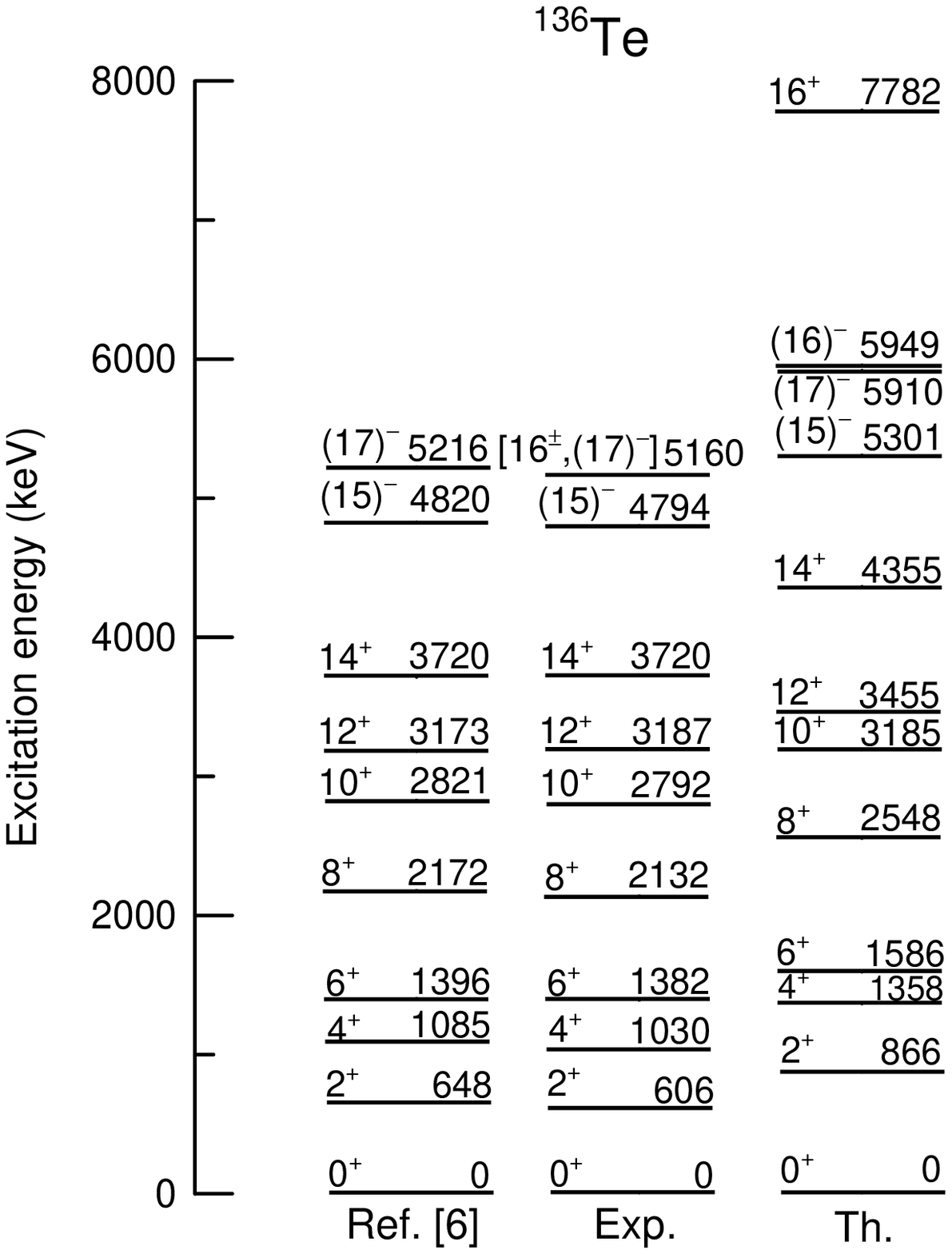}
\caption{Calculated energy levels for positive and negative parity
states of $^{136}$Te in comparison with experiment taken from
Ref.\cite{kor00}.}
\end{figure}

\newpage

\subsection{Transition probabilities}
The electromagnetic transition probabilities $B$($E$2;
0$^+\rightarrow$2$^+$) values calculated for both model spaces and
interactions are compared with those obtained from the measured
lifetimes of states in different nuclei. The radial integral
involved in calculation of $E$2 matrix elements are calculated
with harmonic oscillator radial wave function with
$\hbar\omega$=45$A^{-1/3}$--25$A^{-2/3}$ \cite{Brow88}.

Fig.4 shows the comparison of the calculated results from this work
with the experimental values and with the recent calculations of S.
Sarkar and M. Sarkar \cite{suk04}. The effective charges are taken
as for proton $e_p^{eff}$=1.47 and for neutron $e_n^{eff}$=0.72 also
from Ref.\cite{suk04}. From this figure we can see that our work is
in better agreement with the experiment than those calculated by
Sakar and Sakar at neutron number N=80 and N=82, but their result
for N=84 are better than our work and it seems that their
modifications enhances the prediction of level spectra for
$^{136}$Te as well as the transition rates.

\begin{figure}
\centering
\includegraphics[width=2.5 in]{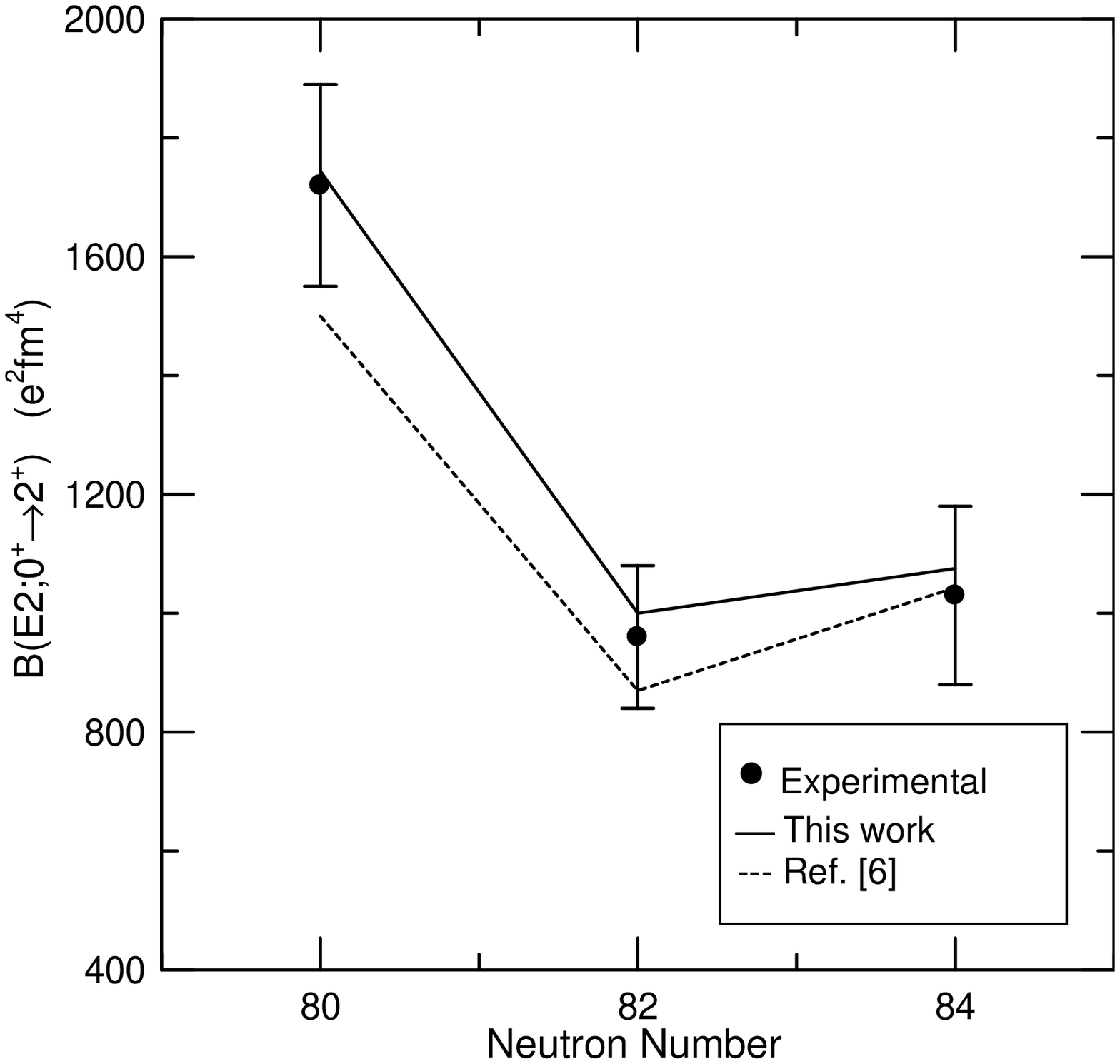}
\caption{Comparison of the calculated $B$($E$2;
0$^+\rightarrow$2$^+$) from this work (solid line) with the
experimental values (closed circles) and with the theoretical work
from Ref.\cite{suk04} (dashed line) for $^{132, 134, 136}$Te .
Experimental data are taken from Refs.\cite{rad02,ter02}. }
\end{figure}
\section{Summary}
Unrestricted large-scale shell model calculations were performed
using the model space SN100PN with hamiltonian SN100PN for $^{132,
134}$Te by choosing the core as $^{100}$Sn rather than $^{132}$Sn
which is taken as core by most authors from previous theoretical
work and a conclusion were drawn that the choice of the core as
$^{100}$Sn does not effect the calculations and this is due to the
fact that the same proton model spaces for both cores were employed
for the calculations .The core $^{132}$Sn were chosen with the
effective interaction KH5082 by choosing suitable single-particle
energies (SP) quoted from Ref. \cite{BA05} for the nucleus
$^{136}$Te. The results of this work are compared with the recently
available experimental data and with the best results achieved from
the previous theoretical work. Overall good agreement were obtained
for all nuclei up to $J \leq 10$, but for high spin states $J > 10$
the model fail to reproduce the experiment. The transition rates are
in excellent agreement with the experimental $B$($E$2;
0$^+\rightarrow$2$^+$) values and and in consistent with the
previous theoretical work.
\section{Acknowledgements}
The author would like to acknowledge the financial support and
warm hospitality from the Abdus Salam International Centre for
Theoretical Physics (ICTP).

\end{document}